\documentclass{iopart}
\usepackage{graphicx,color}
\usepackage[normalem]{ulem} 

\definecolor{red}{rgb}{1,0,0}

\begin{document}

\title{
Evolving cellular automata for diversity generation and pattern recognition:
deterministic versus random strategy
}

\author{Marcio Argollo de Menezes$^1$, Edgardo Brigatti$^2$, Veit Schw\"ammle$^3$}
\address{$^1$ Instituto de F\'{\i}sica, Universidade Federal Fluminense, 
  Campus da Praia Vermelha, 24210-340, Niter\'oi, RJ, Brasil \\
  National Institute of Science and Technology
for Complex Systems - Rua Xavier Sigaud 150, 22290-180 Rio de Janeiro-RJ, Brazil.}
\address{$^2$ Instituto de F\'{\i}sica, Universidade Federal do Rio de Janeiro,  Av. Athos da Silveira Ramos, 149, 21941-972, Rio de Janeiro, RJ, Brazil.}
\address{$^3$ Department of Biochemistry and Molecular Biology, University of Southern Denmark, Campusvej 55, 5230 Odense, Denmark} \ead{veits@bmb.sdu.dk}

\begin{abstract}

Microbiological systems evolve to fulfill their tasks with maximal
efficiency.  The immune system is a remarkable example, where self-non
self distinction is accomplished by means of molecular 
interaction
between self proteins and antigens, triggering affinity-dependent
systemic actions.  
Specificity of this binding and the infinitude of potential antigenic 
patterns call for novel mechanisms to generate antibody diversity.
Inspired by this problem, 
we develop a genetic
algorithm where agents evolve their
strings in the presence of random
antigenic strings and reproduce with affinity-dependent 
rates.
We ask what is the best strategy to generate diversity if
agents can rearrange their strings a finite number of times.
We find that endowing each agent with an inheritable 
cellular automaton rule for performing rearrangements 
makes the system more
efficient in pattern-matching than if transformations are totally random.  
In the former implementation, the population 
evolves to a stationary state where agents with different automata
rules coexist.
\end{abstract}




\date{\today}

\maketitle


\section{Introduction}

Biologically motivated models serve twofold: to give insight on the
possibilities of real processes and systems
\cite{membrane_sim,gillespie,argollo_pnas} or to inspire the
development of new artificial devices, such as neural networks
\cite{pitts,hopfield_model} and DNA computers
\cite{adleman-dnacomputer,shapiro}, touching upon issues as the
definition of life, computation and self-awareness
\cite{whatislife,emergent_comp,immune_model_perelson,bacteria_computer,turing_plant}.
We advance on the latter and seek inspiration in the problem of
diversity generation by the adaptive immune system of vertebrates,
where protein receptors expressed by B cells (called antibodies)
``recognize'' complementary antigenic patterns by means of very
specific molecular interactions that initiate an immune response
whenever a threshold affinity is reached \cite{bcell_affinity,thecell,janeway_book}. 
Each B cell expresses its own unique receptor and a
human can make about $10^{12}$ different receptors
\cite{jerne,thecell,janeway_book}, an astonishing
number if compared to the number of genes in its whole genome 
(about $50,000$) making it impossible to have antibody genes encoded on
DNA. Instead, a relatively small number of disjoint gene segments is
inherited and the antibody region relevant for binding is assembled
during B cell development by rearrangement of some gene segments
\cite{ais-review, janeway_book, biomonitor}. After stimulation by antigen,
B cells reproduce and introduce further mutations on the antibody
binding region greatly increasing diversity \cite{thecell,
  antibody-evolution, antibody-antigen}. 

One might ask whether these genomic modifications are completely
random or if they are guided by some organizing principles
\cite{affinity_nonrandom,affinity_nonrandom1}. Inspired by this
question, we study how one could improve the probability that an
immune system with its antibodies can recognize a random antigen. In
particular, we encode each molecular pattern relevant to binding by
$L$ binary features on Hamming shape space \cite{perelson-bitstring}
and allow for a finite number of modifications of each string
characterizing an antibody.  We ask if cellular automata (CA) rules of
Wolfram type can outperform the rule of random search, when strings
are randomly shuffled. We develop a genetic algorithm where agents
from a population are faced with an antigen's string, perform the
specific computational task of string matching \cite{melanie} and
reproduce if overlap exceeds an arbitrary threshold $T$. Minimum
overlap for reproduction can be achieved by evolving the agent's
string a finite number of times according to its specific grammar
rules.
Algorithms inspired by  human immune system are 
abundant in the literature and,
even if it is not possible to identify one archetypal model 
\cite{greensmith10}, they are usually named
Artificial Immune Systems \cite{forrest93}.
In this contest, our approach 
is built on the fundamental difference
of evolving not simple bit-strings but rules, in the form of 
Cellular Automata \cite{melanie}.

This study has been clearly motivated by
the determinism versus randomness debate
with respect to the diversity generation in the immune systems.
However, our goal is not the definition of a toy
model for simulating  the biological immune system.
In contrast, we focus on replying to the general and abstract 
question whether it is possible to obtain a better efficiency in the pattern 
recognition task using a random or a deterministic computation.
This question is strongly connected with the exploration of how collective 
computation can emerge throughout an evolutionary stochastic process
\cite{emergent_comp, melanie, hordijk}.
A better understanding of this approach could 
generate methods for information processing 
and engineering of new forms of computing systems.\\

The paper is organized as follows. 
In section \ref{sec:random} we study, both analytically and
numerically, the case where the only rule is random shuffling. As
Perelson and Oster,\cite{perelson-bitstring}, we find a steplike
behavior for the probability that an antibody will bind a
random antigen as a function of the minimum overlap for reproduction. 
In section \ref{sec:automata}, we analyse the case where agent
rules are those defining elementary cellular automata and find that
efficiency in the pattern 
recognition task is enhanced. Moreover, maximal
efficiency is achieved when agents with different automata rules
coexist, showing that in this system unsupervised collective
computation emerges from evolution.

 \section{The random model}
 \label{sec:random} 

We develop a genetic algorithm where agents coexist in a population
that, at each time step $t$, faces a recognition challenge originated
from a randomly chosen bit string $Y$ of size $L$ (agents' strings are
of same size) that persists for $P$ time steps before being replaced
by another random string.  One time step is accounted for when all
agents have undergone the following selection rules:

$(1)$ Death with population-dependent rate  $\frac{N(t)}{K}$ where $K$ is
the carrying capacity of the medium and $N(t)$ is the number of agents at time $t$.  
This process is responsible for
limiting the size of the total population.

$(2)$ Overlap-dependent replication: After assembling a random string
$X$, the agent determines its affinity with $Y$ as the Hamming
distance $H(X,Y)$ from antigen $Y$. Replication occurs if $H(X,Y)\le T$
where $L$ is the size of the string.  Step 2 is repeated $S$ times by
each agent and reproduction adds a new agent to the population.\\

The last step of rule $(2)$ mimics the mechanisms of diversity
generation. This model is implemented on a computer, in which an
initial population of $P_0=10^3$ agents with strings of size $L=32$
evolves in a medium with carrying capacity $K=10^5$, eventually
reaching a steady state. In this state an average population is
estimated over $10^5$ time steps 
(simulation parameters are summarized in Table~\ref{table:params}).

We repeat this procedure for
different values of $T$ and $S$ and investigate the effects of binding
specificity and sequence recombination on the average repertoire size,
$N_{S,T}$.
This quantity can be obtained in the mean-field level from the solution of
\begin{equation}
N(t+1)-N(t)=G_S [N(t)-\frac{N(t)^2}{K}]-\frac{N(t)^2}{K}
\end{equation}

when $N(t+1)=N(t)=N_{S,T}$. Here, $G_{S,T}=\sum^{S}_{j=0}
F_T(1-F_T)^j$ is the probability that matching with antigen has
occurred in at most $S$ attempts and $F_T=2^{-32}\sum_{i=0}^{T}
 {32 \choose i}$
is the probability of occurrence of two random strings
with Hamming distance less or equal to $T$.

\begin{table}
\caption{Parameters of the model.}
\begin{center}
\begin{tabular}{l|r}
Parameter & Meaning\\
\hline

K & carrying capacity (controls population size)\\
T & threshold for Hamming distance\\
P & time steps each stimulus remains in the system\\
S & number of tests\\
M & mutation rate of CA
\end{tabular}
\end{center}
\label{table:params}
\end{table}

\begin{figure}
\begin{center}
\includegraphics[angle=270,width=0.8\textwidth]{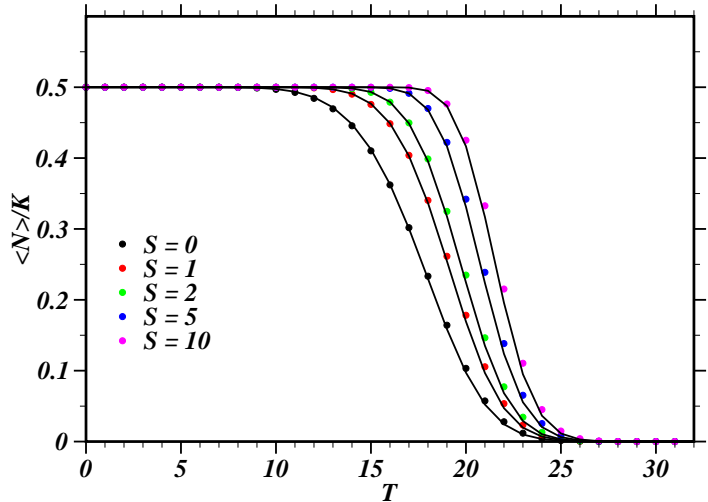}
\end{center}
\caption{The random model and its analytical description match
  perfectly and exhibit a sharp transition in the size of the mean
  population as a particular value of the dissimilarity threshold $T$
  is reached.  Results for different values of $S$ are presented where
  circles depict simulations and solid lines represent the mean field
  description.  Mean populations were averaged over the last $100000$
  time steps after a stationary-like state was reached in the
  simulations (other parameters: $K= 10^5, P=100$).  }
\label{Fig_mpop}
\end{figure}

For  $T\le 10$, $G_0\approx 1$ and so the population at equilibrium is
  equal to $K/2$. For higher $T$ values the population decreases until
  it becomes extinct.  Increasing the $S$ values leads to a higher
  reproduction rate (larger $G_S$ values) which maintains the
  equilibrium population nearer to the classical equilibrium solution
  of $K/2$.  In figure \ref{Fig_mpop} we can appreciate how well the
  mean field description captures the results generated by the
  simulations.
The important global quantity which we need to quantify is 
the success in the recognition task, obtained by evaluating 
the efficiency of the system.
One simple measure of efficiency can be the ratio between the
total number of successful recognitions ($H\le T$) 
and the total number of performed tests.

\begin{figure}
\begin{center}
\includegraphics[angle=270,width=0.8\textwidth]{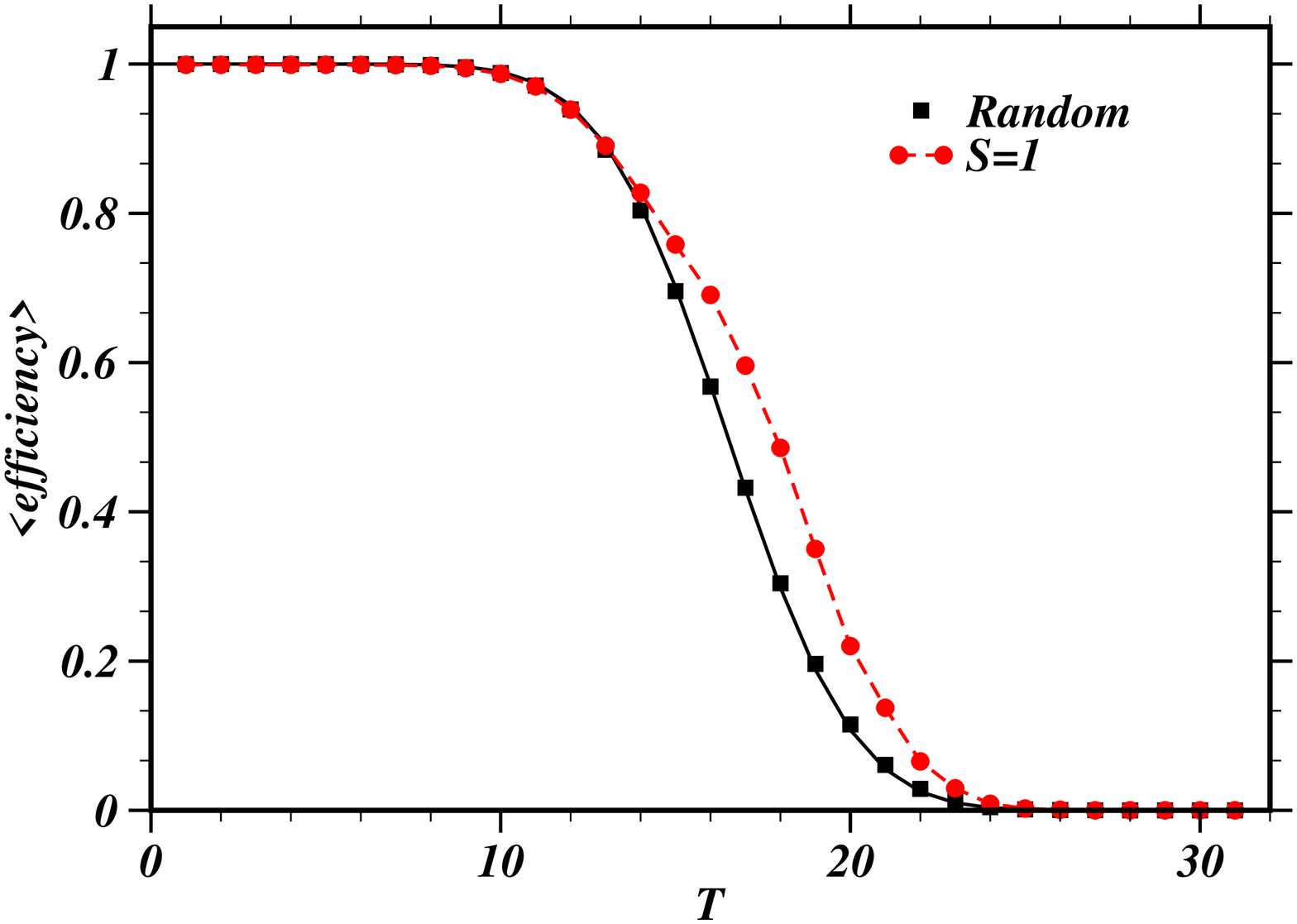}
\includegraphics[angle=270,width=0.8\textwidth]{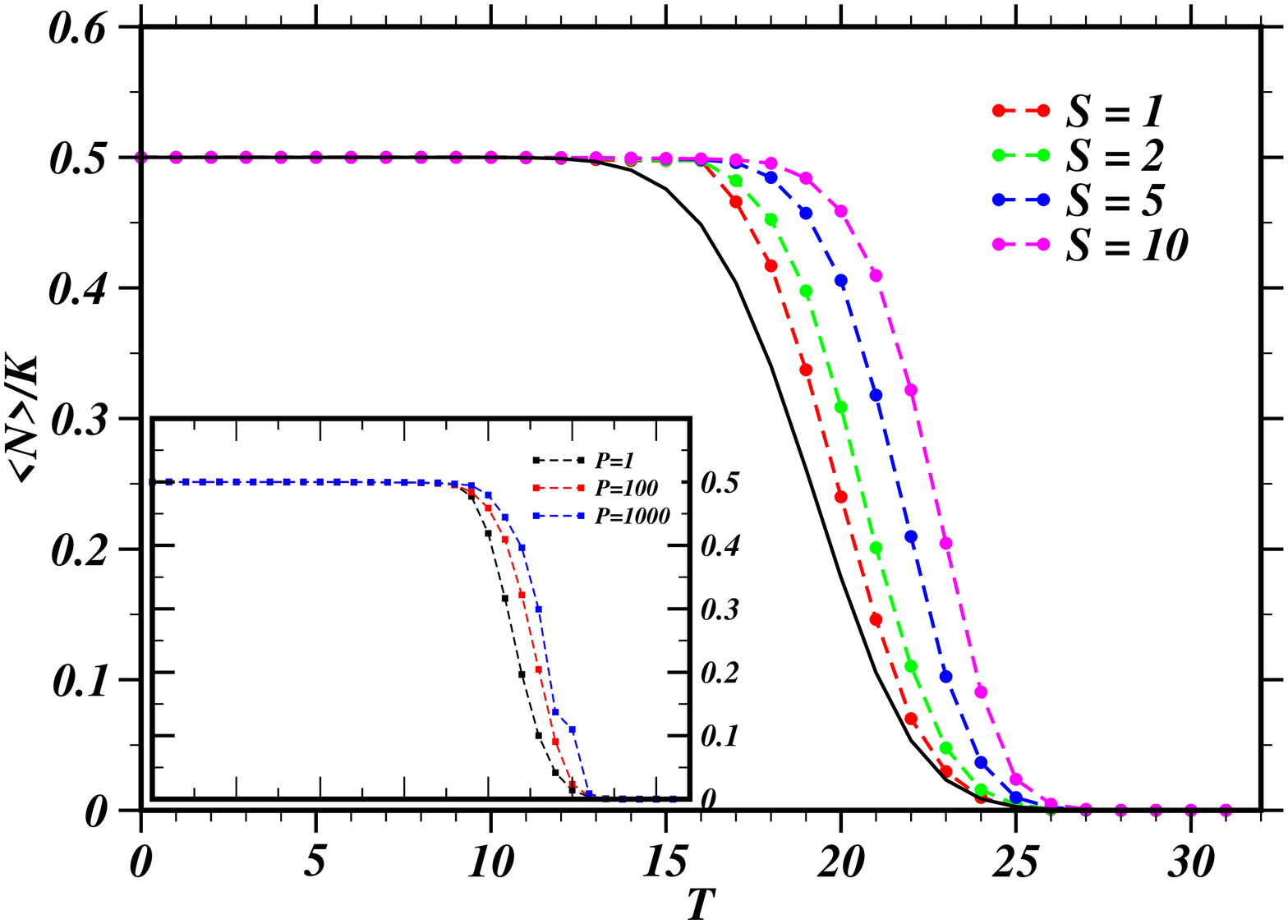}
\end{center}
\caption{\small The CA model outperforms the random model both in terms of population 
size and efficiency. 
Top: Efficiency as a function of the threshold $T$.
The squares represents the data obtained by the model of random somatic mutations. 
The solid line shows the perfect match with the function $F_{T-32}$.
The circles represent the data obtained by the CA model for structured 
somatic mutations with $S=1, M=0.01$.  
As can be appreciate, cellular automata can be more efficient
in bringing antibodies closer to antigens ($K= 10^5, P=1000$).
Bottom: Mean population versus dissimilarity threshold $T$ for different $S$ ($P=100$). For comparison, 
the solid line represents results for the random model when $S=1$.
Inset: mean population versus $T$ for different values of $P$ ($S=10$).
Averages were taken over the last 100000 time steps after a stationary state 
was reached and divided by the the carrying capacity $K$ 
(other parameters: $M=0.01, K= 10^5$).}
\label{Fig_meff}
\end{figure}

Figure \ref{Fig_meff} depicts efficiency in the recognition task
as a function of the threshold $T$ when a stationary regime is
reached. For the random model, efficiency is equal to the 
probability of two 32-bit strings to have a Hamming distance 
less or equal to $32-T$: $F_{32-T}$. 
The efficiency of the model does not depend on the other parameters $S$ and  
$P$. 
As expected, the requirement of larger overlaps between stimulus and agents leads to a sudden decrease of the efficiency.

Our dynamics can be illustrated using an abstract description.
Given a metric space ${\mathcal V}$ , a stimulus is represented by a feature
vector ${\vec x}=(x_0,...,x_N)$.
A agent is represented by a vector ${\vec
  y}=(y_0,...,y_N)$. The distance $|{\vec x} - {\vec y}|$ decides whether a test is successful and the agent reproduces.
Working with binary features, as in our case, the
distance between stimulus ${\vec x}$ and detector ${\vec y}$ can be
given by the Hamming distance $H$ (the number of distinct binary
features).  The agent carries out random jumps of ${\vec x}$
which might move it closer to ${\vec y}$. 
From the analysis developed 
by Perelson and Oster  \cite{perelson-bitstring},
in the continuum limit,
a step-like behavior is found for the probability 
of binding a random antigen (stimulus). 
These results are analogous to the ones we have presented for our
model and, effectively,
 our simulations, for $S=0$, correspond to a discrete version of the 
 model studied in \cite{perelson-bitstring}.

\section{The CA model}
\label{sec:automata}

In the following we introduce the CA model which is motivated by the analogy between antibody generation and grammatical structures.
Each agent is now characterized by one rule to deterministically change its bit-string.
This rule is taken from one of
all possible $256$ elementary Wolfram cellular automata \cite{wolfram}.
These automata are composed by a one-dimensional array of two-state 
cells and by rules operating on the nearest neighbors.

Now, the model is based on the following steps:

1) Each agent dies with a rate  $\frac{N(t)}{K}$
where $K$ is a carrying capacity.

2) Surviving agents get a random bit-string.
They reproduce if a positive presentation, within $S$ tests, is reached.
After each unsuccessful test ($H>T$), the agent's CA 
rule is applied on the bit-string and 
the mutated string is re-compared with the stimulus.
Successful detection generates one new agent with 
the same CA as the ancestor or, with probability $M$, a different random CA rule.
The stimulus bit-string is randomly generated every $P$ time steps.\\

In Figure \ref{Fig_meff}, we show the mean population as a function of the 
threshold $T$ for different values of $S$. For all $S$ values, we notice that
the mean population is larger than the one of the random model for $T$ values where strong selective processes are forcing adaptation of the
CA system. 
In the inset, we present the mean population as a function of the threshold $T$ for different values of $P$. 
It is possible to see how for higher $P$ values the population grows, indicating that 
it reaches an adapted phase with
a structure in the CA rule distribution, which allows more efficient recognition of stimuli.

\begin{figure}[htb]
\begin{center}
\includegraphics[angle=270,width=0.8\textwidth]{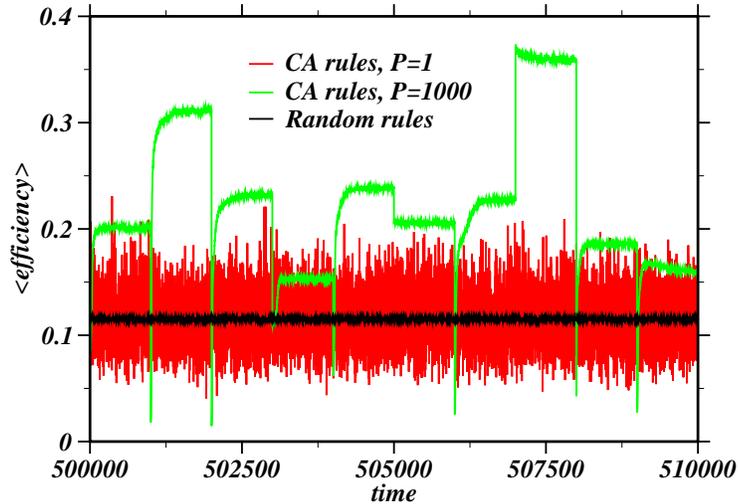}
\end{center}
\caption{\small Fast adaption of the system
in the CA model leads to better performance. 
Data show the temporal behavior of the efficiency. For large enough $P$, the CA model
outperforms the random model.
The green curve is obtained for $P=1000$, the red for $P=1$,
and the black one is the result generated by the random model.
Parameters: $T=20, M=0.01, K=10^5, S=10$. }
\label{Fig_dyn}
\end{figure}

The performance improvement of the CA model can be quantified looking at the efficiency measure. 
As shown in Figure \ref{Fig_meff}, CA rules perform better than random changes. 
These results can be clarified looking at the time evolution of the efficiency
for a given simulation (Figure \ref{Fig_dyn}). 
The efficiency of the CA model is higher than in the random model if the same stimulus is presented for a sufficiently long time.
In fact, if $P=1$, the CA model exhibits the same mean efficiency but higher fluctuations than the random model.
The system is not capable to adapt to the new stimulus within one iteration.
In contrast, a selective dynamics operates for $P=1000$.
After a change of the stimulus, the efficiency drops down, 
followed by a rapid transient where the efficiency grows towards a new plateau higher than the
corresponding value for the random model.
This is because the most efficient CA rules become selected 
and thus initially random agent
can be successfully mutated closer to the stimulus. 
Specific CA rules map specific sub-spaces which contain strings close to the 
stimulus. These rules are able to take different random strings and to
take them closer to the antigen following deterministic paths.   

\begin{figure}[htb]
\begin{center}
\includegraphics[angle=270,width=0.8\textwidth]{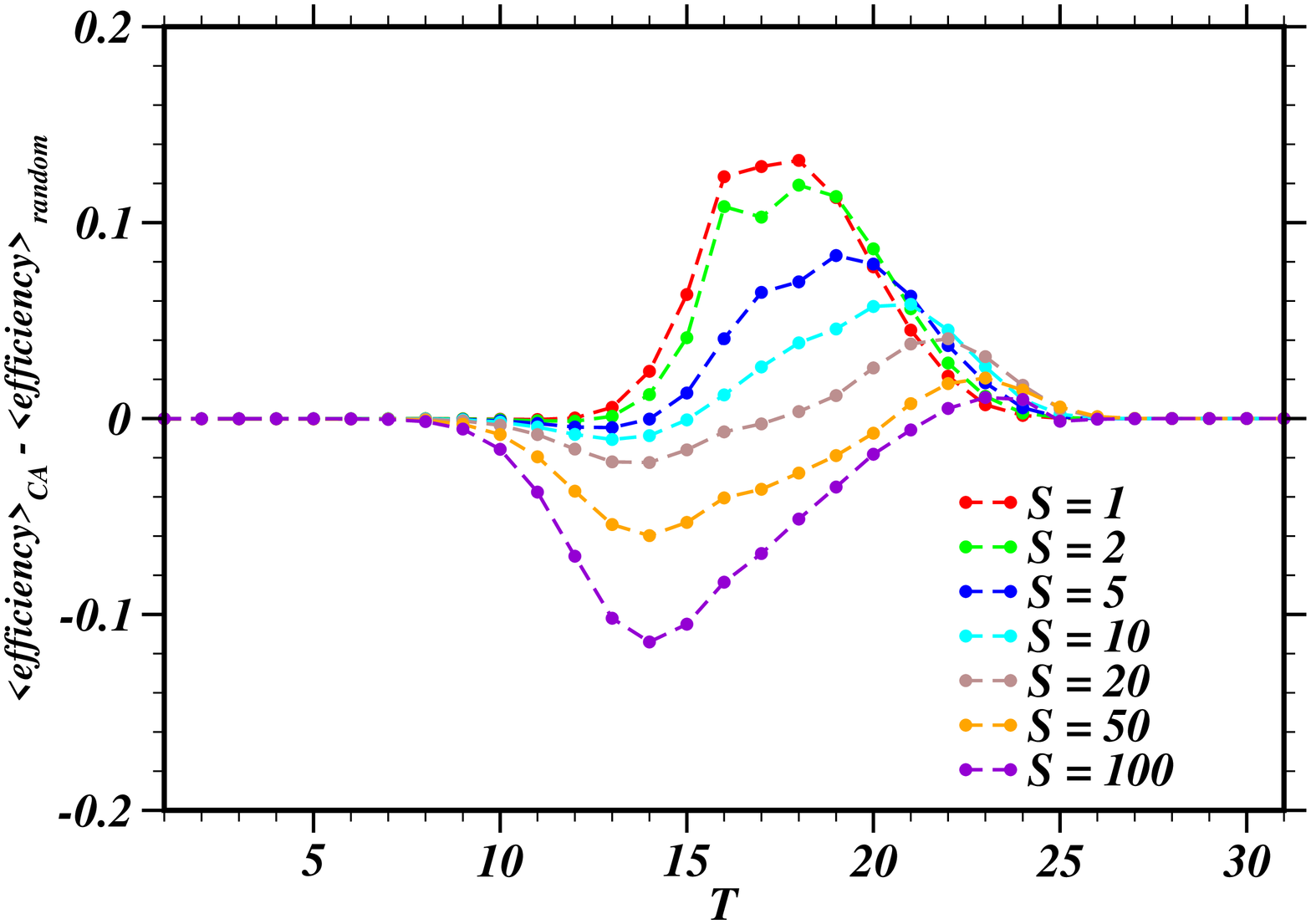}
\end{center}
\caption{\small
The CA model outperforms the random model for small $S$-values. 
The figure shows the difference between the efficiency of the CA model and the one of the random model 
 as a function of the threshold $T$ for different values of $S$. Each point is the time average over 
the last 500000 steps (Parameters: $K= 10^5, M=0.01, P=100$).
}
\label{fig_caeff}
\end{figure}

In Figure \ref{fig_caeff}, we analyzed the efficiency in the recognition task for different values of $S$.
We can see an improvement up to more than 40\% ($S=1, T=18$). 
For high $S$ values, this advantage reduces and for $S>20$ the random model begins to outperform the CA model.
In general, efficiency increases for higher $P$ values, when the selection can effectively operate defining the ensemble of the bests CA rules.

\begin{figure}[htb]
\begin{center}
\includegraphics[angle=270,width=0.8\textwidth]{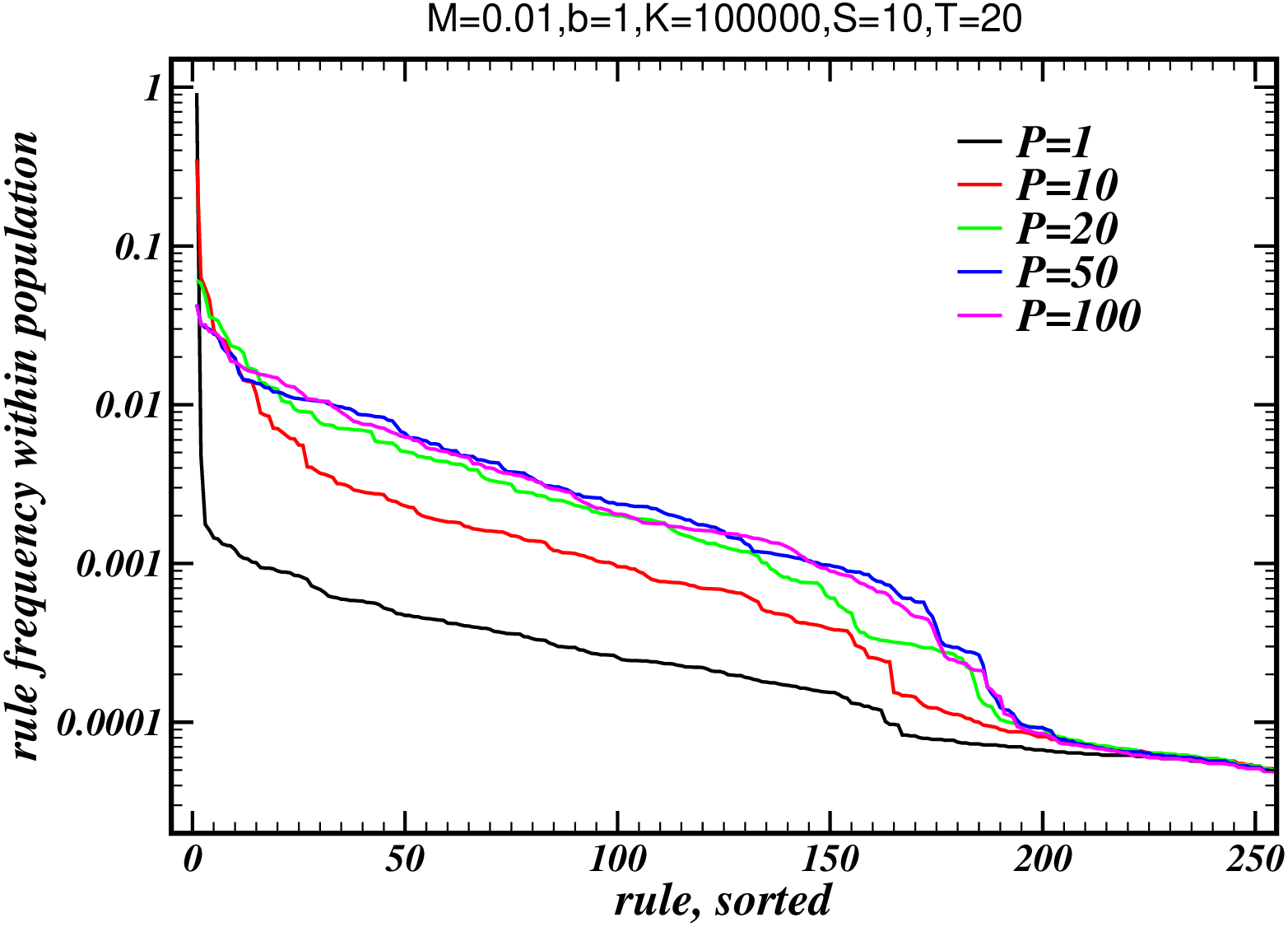}
\includegraphics[angle=270,width=0.8\textwidth]{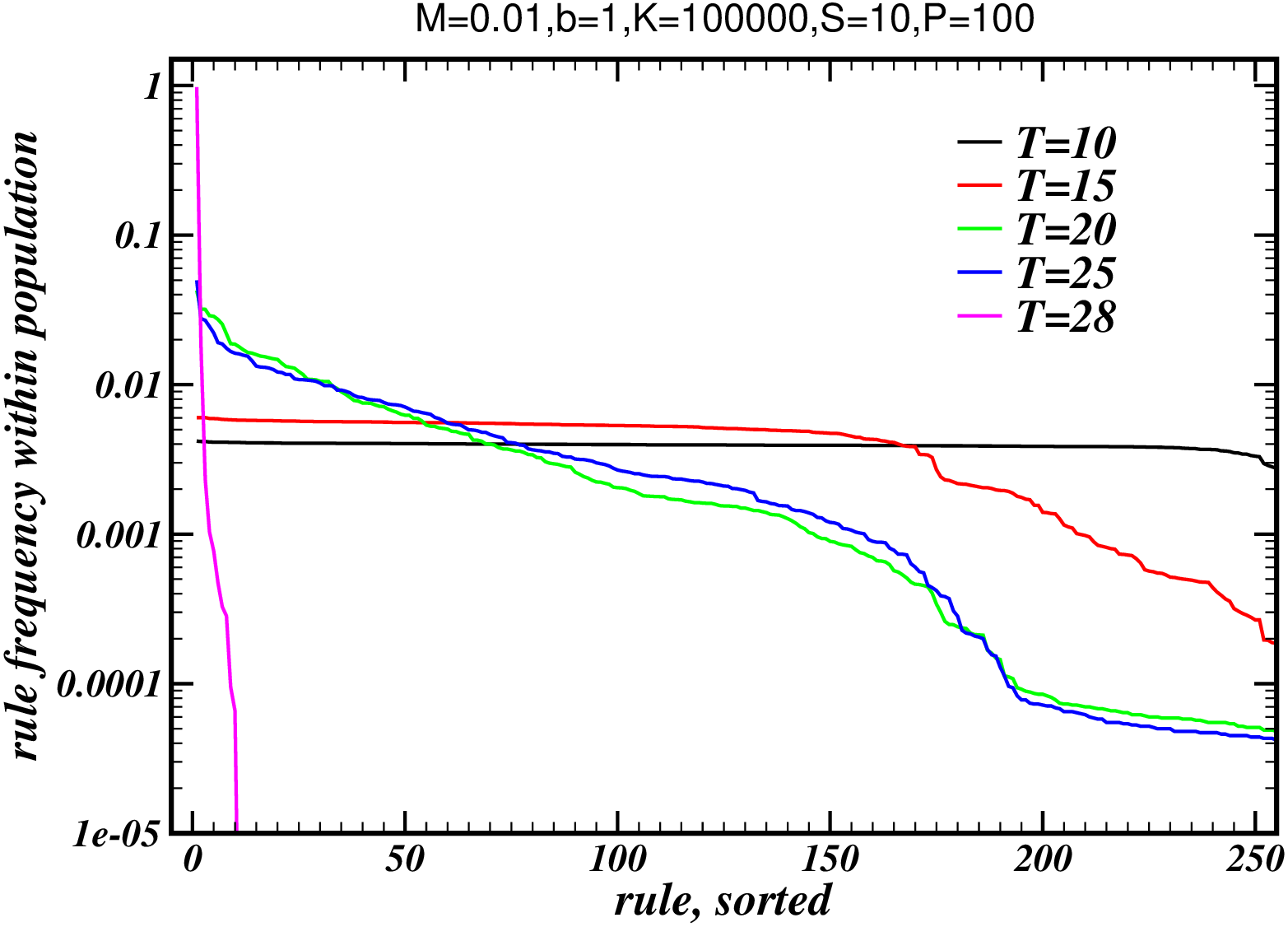}
\end{center}
\caption{Top: Zipf plot for the rules present in the steady state for different $P$ values ($T=20$). 
Botton: Zipf plot for the rules present in the steady state for different $T$ values ($P=100$). 
All data are averaged over the last 500000 time steps ($M=0.01, K=10^5, S=10$).
}
\label{fig_dist}
\end{figure}

Furthermore, we studied the distribution of the population
in terms of the CA rules. Figure \ref{fig_dist} depicts the Zipf plot of the rules.
If reproduction success is not affected by the 
recognition operation ($T\le10$), all the possible CA rules
are maintained in the population generating a flat distribution with equal 
probability for each rule.
In contrast, for higher selection pressures, some rules are selected over the
ensemble of all the CA rules and a structured distribution appears.  
 The rules coexist in the population and they correspond to the ones which 
 allow better performances in the recognition task.
For very high $T$ values, only a minimal fraction of rules survive. 
This happens in correspondence to a very small population where random drift effects become dominant.

\begin{figure}[htb]
\begin{center}
\includegraphics[angle=0,width=0.7\textwidth]{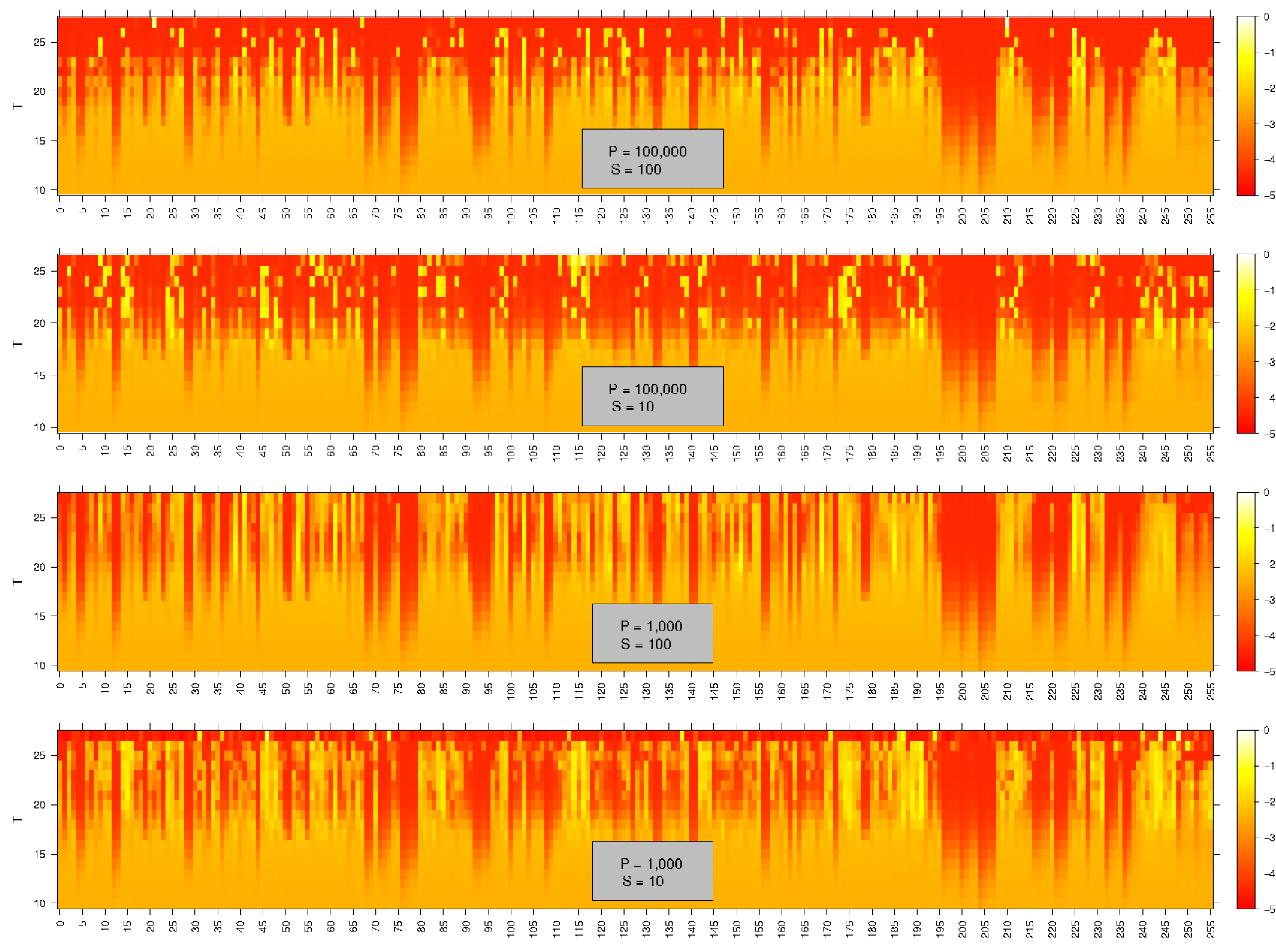}
\end{center}
\caption{Suppression of a subgroup of CA rules for intermediate $T$ values. The frequency of the CA rules ($\log_{10}$-values) is
shown 
as a function of the threshold $T$ (on the ordinate) for different values of $P$ and $S$ (other paramters: $M=0.01$ and $K=10^5$). 
}
\label{fig_heat1}
\end{figure}

Figure \ref{fig_heat1} represents the frequency of the CA rules for different values of $P$ and $T$.
From these figures it is clear how when selection is well operating (hight $P$ and small $S$) a subset of the rules is suppressed and a structured population, with a larger number of lively and coexisting CA rules set on. 
We tried to quantify if this subpopulation of CA rules can be 
related to some particular class following the heuristic Wolfram's classification scheme, 
but unfortunately we were not able to distinguish any specific class of CA rules among 
the ones that better perform in our simulations. 
In contrast, an assortment of CA rules from different classes persists in the population.
In Table \ref{tab_1} we present some of the most successful rules in a specific simulation where $T=19$ and $P=1000$.

\begin{table}
\begin{center}
\begin{tabular}{|c|c|}
\hline
Cellular Automaton & Frequency \\ 
\hline
1  & 0.1935\\	
256 & 0.1256\\	
248       & 0.0842\\
128 & 0.0839\\
192 & 0.0535\\	
252 & 0.043\\
58  & 0.0272\\	
51  & 0.0228\\
52  & 0.0181\\
56  &  0.0169\\
20  & 0.0166\\
19  & 0.0165\\
64  & 0.016\\
168 & 0.0156\\
49  & 0.0153\\
232 & 0.0146\\
244  & 0.0118\\
\hline
\end{tabular}
\end{center}
\label{tab_1}
\caption{ \small Best performing CA, identified by their number, and the corresponding frequency in the population.
Data are averaged over the last 500000 time steps of one simulation 
($M=0.01$, $K=10^5$, $S=1$, $T=19$, $P=1000$). 
}
\label{tabula}
\end{table}

\begin{figure}[htb]
\begin{center}
\includegraphics[angle=0,width=0.7\textwidth]{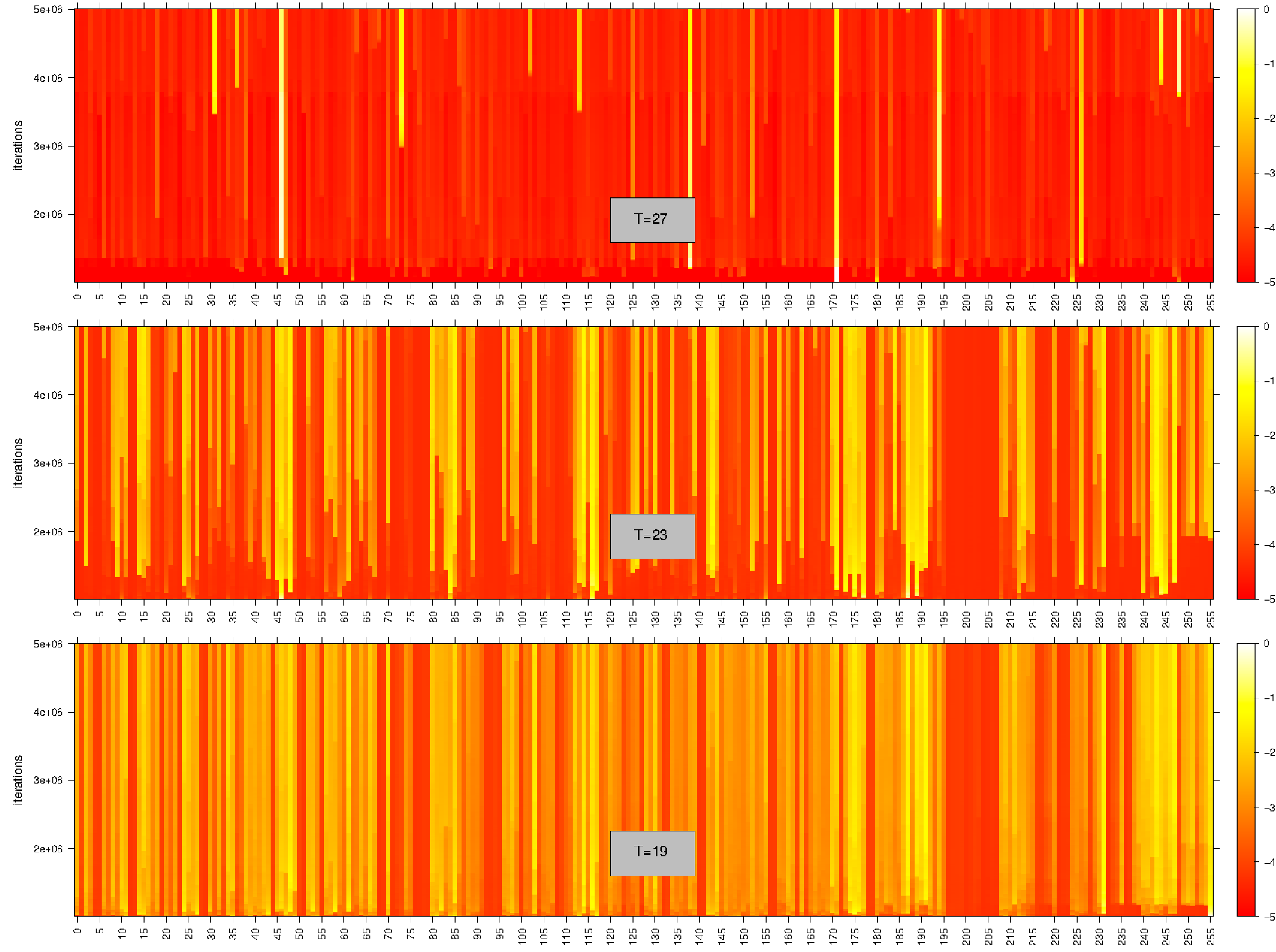}
\end{center}
\caption{\small Fast switching on but slow switching off of new rules when selective pressure is high.
The temporal behavior of the $\log_{10}$-frequencies of the CA rules 
are shown for different $T$ values.
For higher $T$, the system relies on a small number of highly abundant CA rules (parameters: $S=10$, $P=1000$ and $K=10^5$).}
\label{fig_heat2}
\end{figure}

The frequencies of the CA rules exhibit a peculiar temporal behavior in simulations where the selection pressure is high (e.g. $T=27$).
Fig.~\ref{fig_heat2} shows that the introduction of new stimuli leads to a sudden on-switching of single rules that begin to dominate the
population. They however remain in the system over several cycles of new stimuli, leading to long-term correlations between
different stimuli due to memory effects.
This also explains why prevailing CA become selected without following the general trends observed in Fig.~\ref{fig_heat1} for intermediate $T$
values. Rules are selected depending on the context of their co-efficiency with the other rules already in the system. 

As consequence, the CA model relies on the possibility to dispose of a large number of different rules that might be switched on as soon 
their specific properties are required. A reduction to the mostly best performing rules therefore results in a decrease of its performance to
recognize random patterns for high selection pressures (large $T$), or, in other words, demanding recognition tasks.
 
 \section{Conclusions}

We presented the study of a collective model for pattern recognition inspired by the basic biomolecular mechanisms that enable an immune system to detect new antigens.
We explore 
how different mechanisms of antibodies diversity generation
can improve the performance in antigen recognition.
As usual, we represent antigens and antibodies by using bit strings 
and we test two possible strategies for generating antibodies diversity: 
to randomly shuffle  or to apply deterministic rules to the strings which 
represent them.
In the last case we have been influenced by the Jerne's analogy 
between some properties of immunologic system and the concept of
generative grammars \cite{jerne}. 
We have implemented these ideas introducing a genetic algorithm which evolves 
an ensemble of Wolfram's cellular automata
which performs the computational task of string identification.  
Thanks to the employment of  evolutionary simulations 
based on a genetic algorithm, we find that not one, but a group of rules, 
performs the recognition task better than dull random shuffling.\\

Our study outline interesting results which can be useful 
for general information processing.
Because of the biomolecular nature of the biological 
problem which we have theoretically explored, 
we speculate that our abstract result could be transposed 
into practical applications for designing 
computational devices for pattern recognition implemented 
by the means of a biomolecular computer.

\ack
MAM acknowledges partial financial support
from the Brazilian agency Conselho Nacional de Desenvolvimanto
Cient\'{\i}fico e Tecnol\'ogico (CNPq). VS was supported by the Danish
Council for Independent Research, Natural Sciences (FNU).\\

\bibliographystyle{unsrt}
\bibliography{automata2011}

\end{document}